\documentclass[a4paper,11pt]{article}
\usepackage{pos}

\title{Renormalization of non-local gluon operators in lattice perturbation theory}

\author*{Demetrianos Gavriel}
\author{Haralambos Panagopoulos}
\author{Gregoris Spanoudes}

\affiliation{University of Cyprus,\\
  1 Panepistimiou Avenue, 2109 Aglantzia, Cyprus}

\emailAdd{gavriel.demetrianos@ucy.ac.cy}
\emailAdd{panagopoulos.haris@ucy.ac.cy}
\emailAdd{spanoudes.gregoris@ucy.ac.cy}

\abstract{In this study, we investigate the renormalization of a complete set of gauge-invariant non-local gluon operators up to one-loop in lattice perturbation theory. Our computations have been performed in both dimensional and lattice regularizations, using the Wilson gluon action, leading to the renormalization functions in the modified Minimal Subtraction $(\overline{\text{MS}})$ scheme, as well as conversion factors from the modified regularization invariant $(RI')$ scheme to $\overline{\text{MS}}$.
\begin{center}
\includegraphics[scale=0.45]{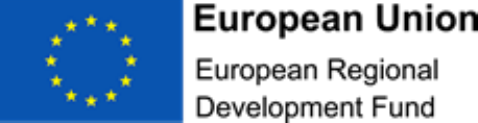}
\includegraphics[scale=0.45]{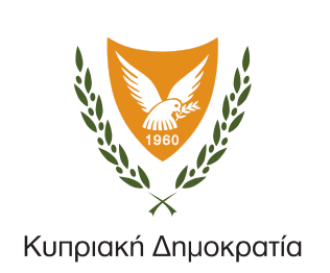}
\includegraphics[scale=0.45]{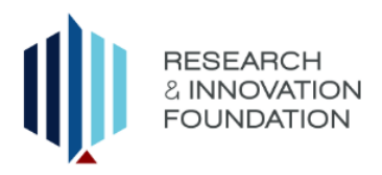}
\includegraphics[scale=0.45]{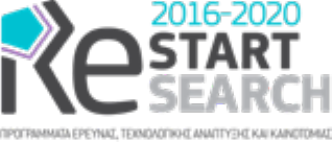}
\end{center}
}

\FullConference{The 40th International Symposium on Lattice Field Theory (Lattice 2023)\\
July 31st - August 4th, 2023\\
Fermi National Accelerator Laboratory\\}

\begin{document}
\maketitle

\section{Introduction}

Parton distribution functions (PDFs) provide insights into the internal composition of hadrons, involving quarks and gluons. PDFs quantify the probability of finding partons with a specific momentum fraction and spin in the infinite momentum frame. They are defined as expectation values of light-cone correlation functions in hadronic states, making it unfeasible to directly compute them on a Euclidean lattice through lattice QCD calculations. 

However, over the last decade, several methods have contributed to significant progress in the direct calculation of PDFs through lattice QCD. For a recent summary of these approaches, refer to \cite{Constantinou2020}. One noteworthy and commonly used approach is the quasi-distribution  method, which utilizes the large momentum effective theory (LaMET) \cite{Ji2013,Ji2014}. Instead of computing light-cone correlation functions, this method calculates quasi-PDFs, which are defined as the matrix elements of momentum-boosted hadrons coupled to gauge-invariant non-local operators that include a finite-length Wilson line. This quasi-observable, which is usually hadron-momentum-dependent but time-independent, can be calculated on the lattice and then renormalized nonperturbatively in an appropriate scheme. Finally, the renormalized quasi-PDF is matched to the PDF through a factorization formula \cite{Ma2014}.

The gluonic contribution of hadron composition has been overlooked compared to quark counterparts, despite playing a crucial role in various physical quantities. Phenomenological data suggest that gluons contribute approximately 40\% of the hadron's momentum at a scale of 6.25 $\rm GeV^2$ \cite{Alekhin2014}. Accurate calculations of the gluon dependent quantities are essential for $J/\psi$ photo production, cross-section of Higgs boson production and jet production, as well as for providing theoretical input to the upcoming Electron-Ion Collider 
.

A crucial aspect in the direct calculation of PDFs from lattice QCD is the nonperturbative renormalization of the quasi-PDFs. As regards quark quasi-PDFs, in Ref. \cite{Constantinou2017} two important features of the Wilson-line operator matrix elements were revealed on the lattice: linear divergences in addition to logarithmic divergences, and mixing among certain subsets of the original operators during renormalization. Efforts to eliminate these linear divergences have been made using various methods, however a complete nonperturbative renormalization program has only recently been developed. Similar effects are expected to be present in the renormalization of non-local gluon operators as well. A recent study \cite{Zhang2018}, using the auxiliary field approach, showed that different components of non-local gluon operators have nontrivial renormalization patterns, making it challenging to evaluate accurately gluon quasi-PDFs. In addition, four gluon operators have been identified as multiplicatively renormalizable, making them suitable for defining some of the gluon quasi-PDFs. 

In these proceedings, we outline the methodology for computing the renormalization of gauge-invariant non-local gluon operators in lattice perturbation theory. We calculate the renormalization functions of non-local gluon operators in the $\overline{\text{MS}}$ scheme to one loop, employing dimensional and lattice regularization, ultimately resulting in the derivation of the relevant mixing matrices. We also compute the conversion factors which are necessary in relating the renormalization functions stemming from a non-perturbative lattice scheme to $\overline{\text{MS}}$.

\section{Formulation}

\subsection{Lattice action}

We consider a non-abelian gauge theory of $SU(N_c)$ group and $N_f$ fermions. To simplify our calculations, we employ the Wilson plaquette gauge action for gluons:
\begin{eqnarray}
\hspace{-1cm}
S_G=\frac{2}{g_0^2} \; \sum_{\rm plaq.} {\rm Re\,Tr\,}\{1-U_{\rm plaq.}\}
\label{eq:gluon_action}
\end{eqnarray}
where
\begin{equation}
    U_{\rm plaq.}=U_\mu(x)U_\nu(x+\hat{\mu})U^\dagger_\mu(x+\hat{\nu})U^\dagger_\nu(x)
\end{equation}

We expect that improved gauge actions, such as the Symanzik improved action, or the implementation of stout-smeared links, will not have an impact on determining the mixing pattern under renormalization of the non-local operators.

\subsection{Definition of operators}

The non-local gluon operators under study are defined in the fundamental representation as:
\begin{equation}
    O_{\mu \nu \rho \sigma} (z \hat{\tau},0) \equiv 2 \ Tr \bigg( F_{\mu \nu}(z \hat{\tau}) W(z \hat{\tau},0) F_{\rho \sigma}(0) W(0, z \hat{\tau})  \bigg)
    \label{eq:nonlocal_operator}
\end{equation}
where $F_{\mu \nu}$ is the gluon field strength tensor and $W(x,x+z \hat{\tau})$ is the path-ordered ($\mathcal{P}$) straight Wilson line inserted for gauge invariance: 
\begin{equation}
   W(x,x+z \hat{\tau}) = \mathcal{P} \  e^{i g_0 \int_0 ^z A_\mu (x+\zeta \hat{\tau}) d\zeta}
\end{equation}
Without loss of generality, the Wilson line is
chosen to lie along the $z$ direction. 

Due to anti-symmetry of $F_{\mu\nu}$, for a fixed choice of the Wilson line there are 36 non-local operators in total by selecting the indices of $O_{\mu \nu \rho \sigma}$ to be in any direction. However, only gluon operators that exhibit multiplicative renormalizability are appropriate for defining the gluon quasi-PDF \cite{Ma2014}. Suitable candidates for the unpolarized gluon quasi-PDF can be provided by \cite{Wang2019}:
\begin{equation}
    \label{eq:gluon_qPDF}
    {\tilde f}_{g/H}^{(n)}(x, P^z) = \mathcal{N}^{(n)} \int \frac{dz}{2\pi  x  P^z} e^{iz x P^z}  \langle H(P)| O^{(n)}(z, 0) |H(P)\rangle
\end{equation}
where $\mathcal{N}^{(n)}$ is a renormalization factor, $x$ is the longitudinal momentum fraction carried by the gluon, $P^\mu=(P^0,0,0,P^z)$ is the hadron momentum, and $H(P)$ is the momentum boosted hadron states. Potential candidates for the gluon operator are denoted here as $O^{(n)}(z, 0)$.
 
\section{Perturbative Calculation - Results}

\begin{figure}
    \centering
    \includegraphics[width=0.8\textwidth]{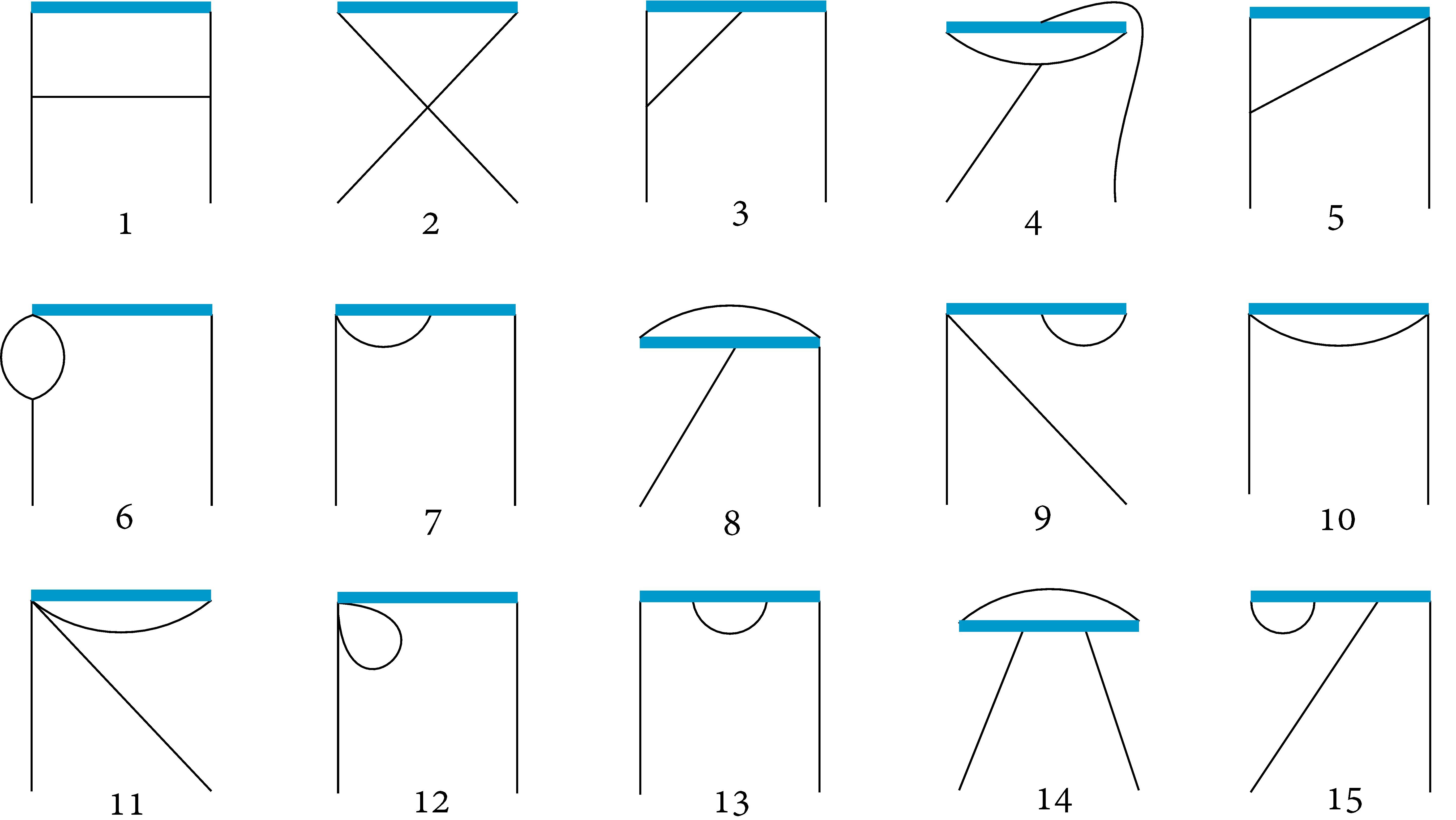}
    \caption{Feynman diagrams contributing to the one-loop calculation of the Green’s functions of the non-local operators. Solid lines represent gluons. The operator insertion is denoted by a solid box.}
    \label{fig:Feynman_diagrams}
\end{figure}

To study the renormalization of the non-local gluon operators, we choose, for convenience, to calculate the following Green's functions:
\begin{equation}
    \Lambda_O = \langle A^{\alpha_1}_{\nu_1} (q_1) \ A^{\alpha_2}_{\nu_2} (q_2) \ O_{\mu \nu \rho \sigma}   \rangle
    \label{eq:Green_f_nonlocal}
\end{equation}
where operator $O_{\mu \nu \rho \sigma}$ is defined by Eq. (\ref{eq:nonlocal_operator}), and $A^{\alpha_1}_{\nu_1} (q_1)\text{,} \ A^{\alpha_2}_{\nu_2} (q_2)$ are two external gluons. The main task in this work is the evaluation of the above bare amputated Green’s functions up to one-loop using dimensional and lattice regularization in the the modified minimal subtraction ($\overline{\text{MS}}$) scheme. The Feynman diagrams of $\Lambda_O^{\text{1-loop}}$ are shown in Fig. (\ref{fig:Feynman_diagrams}).

We define the renormalization factors which relate the bare quantities with the renormalized counterparts through:
\begin{equation}
  O_{\mu \nu \rho \sigma}^R=Z^{-1}_{O_{\mu \nu \rho \sigma}} O_{\mu \nu \rho \sigma} \;\;,\;\; A^R_{\mu} = Z_A^{-1/2} A_{\mu}  
\end{equation}
where $A_{\mu}$ ($A^R_{\mu}$) is the bare (renormalized) gluon field. In the presence of operator mixing, the relationship between the bare and the renormalized operators is appropriately generalised.

The corresponding renormalized amputated Green's functions are expressed as (for simplicity, we omit Lorentz indices whenever they can be understood from the context):
\begin{equation}
    \Lambda^R_O = Z_A Z_O^{-1} \Lambda_O
    \label{eq:Green_f_renormalized}
\end{equation}

\subsection{Dimensional Regularization}

The computations are performed in $D$-dimensional Euclidean spacetime, where $D=4-2\epsilon$, and $\epsilon$ is the regularization parameter. In contrast to 2-point Green's functions involving local operators, the integration results become significantly more complicated due to the presence of both the external momentum and the length of the Wilson line in the integrands. Additionally, there is a nontrivial dependence on the preferred direction of the Wilson line, leading to further complexity. We apply new techniques,  similar to \cite{Spanoudes2018}, for assessing one-loop tensor integrals with an exponential factor in $D$ dimensions. For the elimination of the poles in $\epsilon$, we adopt the $\overline{\text{MS}}$ scheme. 

\subsubsection{Renormalizations Functions}

We start by considering the amputated tree-level Green’s functions, as presented in Eq. (\ref{eq:Green_f_nonlocal}), which read:
\begin{equation}
    \begin{aligned}
        \Lambda_O^{\text{tree}}= \delta^{\alpha_1 \alpha_2} 
        \big( 
        &+ q_{1 \mu} q_{2 \rho} \ \delta_{\nu_1 \nu} \delta_{\nu_2 \sigma} \ e^{-iz q_1 \hat{\tau}}
        + q_{1\mu} q_{2\rho} \ \delta_{\nu_1 \sigma} 
        \delta_{\nu_2 \nu} \ e^{iz q_1 \hat{\tau}}\\
        &- q_{1\nu} q_{2\rho} \ \delta_{\nu_1 \mu} \delta_{\nu_2 \sigma} \ e^{-iz q_1 \hat{\tau}}
        - q_{1\nu} q_{2\rho} \ \delta_{\nu_1 \sigma} \delta_{\nu_2 \mu} \ e^{iz q_1 \hat{\tau}}\\
        &- q_{1\mu} q_{2\sigma} \ \delta_{\nu_1 \nu} \delta_{\nu_2 \rho} \ e^{-iz q_1 \hat{\tau}}
        - q_{1\mu} q_{2\sigma} \ \delta_{\nu_1 \rho} \delta_{\nu_2 \nu} \ e^{iz q_1 \hat{\tau}}\\
        &+ q_{1\nu} q_{2\sigma} \ \delta_{\nu_1 \mu} \delta_{\nu_2 \rho} \ e^{-iz q_1 \hat{\tau}}
        + q_{1\nu} q_{2\sigma} \ \delta_{\nu_1 \rho} \delta_{\nu_2 \mu} \ e^{iz q_1 \hat{\tau}}
        \big)
    \end{aligned}
    \label{eq:Tree_level_green_function}
\end{equation} 
where $z$ and $\hat{\tau}$ is the length and direction of the Wilson line. Notice that the above expression is antisymmetric in $\{\mu , \nu\}$ and $\{\rho , \sigma\}$ as expected. 

Subsequently, we proceed to the 1-loop calculations. We ﬁnd that only diagrams 3, 6 and 13, contribute to the $1/\epsilon$ terms, and therefore the renormalization function of the operators in $\overline{\text{MS}}$ is not affected by the remaining diagrams. However, they contribute to the renormalized Green’s functions. Below we present the $\mathcal{O}(1/\epsilon)$ contributions of the perturbative calculation:
\begin{equation}
    \Lambda_O^{\text{1-loop}}\vert_{\mathcal{O}(1/\epsilon)}= \frac{g^2 N_c}{16 \epsilon \pi^2}
    \left( \delta_{\mu \hat{\tau}}+\delta_{\nu \hat{\tau}}+\delta_{\rho \hat{\tau}}+\delta_{\sigma \hat{\tau}}- \frac{1}{2} \beta \right) \Lambda_O^{\text{tree}}
    \label{eq:oneloop_nonlocal}
\end{equation} 
where $\beta$, the gauge fixing parameter, is deﬁned such that $\beta=0(1)$ corresponds to the Feynman (Landau) gauge. The computation was carried out in an arbitrary covariant gauge, allowing for a direct verification of the gauge invariance of the renormalization functions. It should be noted that, at the one-loop level in dimensional regularization (DR), the pole terms are proportional to the tree-level values for each one of the operators, indicating no mixing with operators of equal or lower dimension. 

The renormalization factor of the gluon field in DR is given by:
\begin{equation}
    Z_A^{\text{DR}, \overline{\text{MS}}}= 1+ \frac{g^2}{16 \epsilon \pi^2} 
    \left( \frac{13 N_c}{6} - \frac{N_c}{2}\left(1-\beta\right) - \frac{2}{3} N_f  \right)
    \label{eq:renorm_factor_gluon}
\end{equation} 
Using the $\overline{\text{MS}}$ condition and Eqs. (\ref{eq:Green_f_renormalized}), (\ref{eq:oneloop_nonlocal}), and (\ref{eq:renorm_factor_gluon}), we find the renormalization function of the operators:
\begin{equation}
    Z_O^{\text{DR}, \overline{\text{MS}}}= 1+ \frac{g^2}{16 \epsilon \pi^2} \left( \left( \frac{5}{3} + \delta_{\mu \hat{\tau}}+\delta_{\nu \hat{\tau}}+\delta_{\rho \hat{\tau}}+\delta_{\sigma \hat{\tau}}\right) N_c- \frac{2}{3} N_f  \right) 
    \label{eq:renorm_factor_operator}
\end{equation} 
We observe that this result depends on the choice of indices for the operators, specifically whether they align with the direction of the Wilson line or not.

As expected from gauge invariance in $\overline{\text{MS}}$, the $\beta$ dependence disappears in the renormalization function of the operators, upon taking into account the gluon field renormalization function. Gauge invariance cannot be ensured in all schemes due to the presence of gauge-dependent renormalized external fields in the Green’s functions.

It is worth mentioning that the renormalization function of the operators is independent of the length of the Wilson line ($z$). There is no dimensionless factor dependent on $z$ that could emerge in the pole part, because the leading pole at each loop cannot depend on external momenta or the renormalization scale. Consequently, $z$ independence is expected to persist at all orders in perturbation theory.

The $\overline{\text{MS}}$-renormalized Green's functions, which equal the finite part of $\Lambda_O$, are very lengthy expressions involving integrals over Bessel functions; their explicit expression will appear in our forthcoming \cite{Gavriel2024}. They are the fundamental ingredient for the construction of the conversion factors between non-perturbatively applicable renormalization schemes and $\overline{\text{MS}}$.

\subsection{Lattice Regularization}

We now focus on computing the bare Green's functions, as given by Eq. (\ref{eq:Green_f_nonlocal}), using lattice regularization. The tree-level Green's functions yield the same result as in dimensional regularization, shown in Eq. (\ref{eq:Tree_level_green_function}).

The 1-loop computation is considerably more complicated than in dimensional regularization due to the subtleties involved in extracting divergences from lattice integrals. To begin, we write the lattice expressions in the form of a sum of continuum integrals and additional lattice corrections. Noteworthy, these additional terms, although they have a simple quadratic dependence on the external momentum $q$, are expected to have a nontrivial dependence on $z$ as seen in non-local fermion operators \cite{Constantinou2017}. 

Several diagrams (1, 2, 4, 5, 8, 10, and 14) give precisely the same contributions as in DR. This aligns with expectations, considering that these contributions are finite as $\epsilon \rightarrow 0$. Consequently, the limit $a \rightarrow 0$ can be applied right from the beginning, without inducing any lattice corrections. However, we must ensure that we eliminate the overall factor of $1/a^2$, attributed to the presence of the external gluons in the Green's functions, by extracting two powers of the external momentum, $(aq)$.

As an example of the ensuing expressions, we present the one-loop lattice result for diagram 13, which is particularly simple:
\begin{equation}
    \Lambda_O^{\text{d13}} = \frac{g^2 N_c}{16 \pi^2} \left( c_1 + c_2 \, \beta - c_3 \frac{|z|}{a} + 8 \log{\frac{|z|}{a}}\, (2+\beta) \right)  \Lambda_O^{\text{tree}}
    \label{eq:lattice_d13}
\end{equation}
where $c_1=32.24812(2)$, $c_2=14.24059(4)$, and $c_3=79.81936(8)$. Note here the presence of both linear divergence and logarithmic divergence in $a$, features revealed in the non-local fermion operators as well. 

The non-local gluon operators exhibit a rotational symmetry of the octahedral point group with respect to the three transverse directions to the preferred direction of the Wilson line. Considering also parity arguments, there is no requirement of mixing among operators that do not share the same transformations under these symmetries. In particular, we find that mixing is possible only between pairs of operators. An example of such a pair is represented by $\left[ \frac{1}{2} \left( O_{z1z2} + O_{z2z1} \right) , \frac{1}{2} \left(O_{4142} + O_{4241} \right) \right]$, where the transverse directions to the Wilson line, $z$ direction, are denoted by $1, 2, $ and $4$.

Our findings of the possible mixing pairs due to symmetries, along with the one-loop bare and renormalized Green’s functions in lattice regularization will be included in \cite{Gavriel2024}.

\section{Summary}

In this work, we obtain the renormalization functions of non-local gluon operators in the $\overline{\text{MS}}$ scheme using both dimensional and lattice regularizations. In addition, using the one-loop result for the renormalized functions of the operators, we find the conversion factors between the $RI'$ and $\overline{\text{MS}}$ scheme. This will enable the conversion of the corresponding lattice non-perturbative results to the $\overline{\text{MS}}$ scheme. Furthermore, we find that cubic and parity symmetry allow a finite mixing among operators in pairs. These results are relevant for the non-perturbative studies of the gluon quasi-PDF calculation.

\acknowledgments

D.G., H.P. and G.S. acknowledge financial support from the project “Lattice Studies of Strongly Coupled Gauge Theories: Renormalization and Phase Transitions”, funded by the European Regional Development Fund and the Republic of Cyprus through the Research and Innovation Foundation (Project:  EXCELLENCE/0421/0025).



\bibliographystyle{JHEP}
\bibliography{references}

\end{document}